\documentstyle[12pt,epsfig]{article} 
\def \LSP{\widetilde{N}_1}
\def \N2{\widetilde{N}_2}
\def \CH{\widetilde{\chi}^{\pm}}
\def \W1{\widetilde{\chi}_1^{\pm}}
\def \WC2{\widetilde{\chi}_2^{\pm}}
\def \SNU{\tilde{\nu}}

\def \SELR{\tilde{l_R}}
\def \SELL{\tilde{l_L}}
\def \SQ{\tilde{q}}
\def \GLU{\tilde{g}}
\def \MLSP{M_{\widetilde{N}_1}}
\def \MN2{M_{\widetilde{N}_2}}
\def \MCH1{M_{\widetilde{\chi}_1^{\pm}}}
\def \M2CH{M_{\widetilde{\chi}_2^{\pm}}}
\def \MSNU{m_{\tilde{\nu}}}
\def \MSELL{m_{\tilde{l}_L}}
\def \MSELR{m_{\tilde{l}_R}}
\def \GLUM{m_{\tilde{g}}}
\def \MSQ{m_{\tilde{q}}}
\def \ETMIS{\not \!\!  E_T}
\begin{document}
\setcounter{page}{0}
\thispagestyle{empty}

\begin{flushright}
TIFR/TH/96-46\\
TIFR/HECR/DZERO/96-1\\
IC/96/136\\
hep-ph/9609413
\end{flushright}

\begin{center}
{\LARGE\bf  Squark Gluino Mass Limits Revisited For 
Nonuniversal Scalar Masses\\}
\bigskip
{\bf Amitava Datta}\\
{\footnotesize 
International Centre for Theoretical Physics, Trieste, Italy.\footnote{
Permanent Address:
Department of Physics, Jadavpur University, Calcutta 700 032, India.}\\}
{\bf Monoranjan Guchait\footnote{email:guchait@theory.tifr.res.in}\\}
{\footnotesize 
Theoretical Physics Group, Tata Institute of Fundamental Research, \\ 
Homi Bhabha Road, Mumbai 400 005, India.} \\ 

{\bf Nirmalya Parua\footnote{email:parua@tifrvax.tifr.res.in }\\}
{\footnotesize 
High Energy Cosmic Ray Group, Tata Institute of Fundamental Research, \\ 
Homi Bhabha Road, Mumbai 400 005, India.} \\ 
\end{center}
\vskip 5pt

\begin{center}
{\large\bf ABSTRACT} 
\end{center}
\footnotesize
It is shown that if the sneutrino is the second lightest SUSY particle,
then the decay products of squarks and gluinos produced at the TEVATRON collider tend to have i) more leptons, ii) smaller number of jets and
iii) two or more carriers of $\ETMIS$. This may relax the existing limits on the squark and gluino masses. This effect is likely to be even 
more striking as these limits improve with accumulation of data. Numerical
results for signal cross sections are presented and compared with the ones
obtained without a light sneutrino. The possibilities of accommodating this scenario in models motivated by N = 1 SUGRA are discussed.
\normalsize\rm

\newpage
\section{Introduction}

Supersymmetry (SUSY) \cite{R1} is one of the most important 
alternatives to the Standard Model (SM) with  an elegant
solution of the notorious naturalness problem,  provided the masses
of the superpartners are of the order of 1 TeV or less. The search
for SUSY at the TeV scale is, therefore,  a high-priority programme
of current high energy physics. Extensive searches for SUSY at the
present high energy accelerators including the Fermilab Tevatron \cite{R2},
 LEP-1 \cite{R3}  and LEP-1.5 \cite {R4} have yielded negative results and have eliminated certain regions of the parameter space of  the Minimal Supersymmetric extension of the Standard Model (MSSM)\cite{R1}.

However, the parameter space of SUSY is so complicated that the
 limits on the sparticle masses obtained from these  searches
almost always involve certain simplifying assumptions. Past experiences
reveal that such limits often require revisions once such assumptions
are removed. As an example let us consider the early limits on
squark( $\SQ$ ) and gluino ( $\GLU$ ) masses ( $\MSQ$ and $\GLUM$ ) obtained
 by the UA1 \cite{R5}, UA2 \cite{R6} and CDF \cite{R7} collaborations 
 in the jets + missing transverse energy ( $\ETMIS$ ) channel. These 
were based on the assumption  that these sparticles 
directly decay, with 100$\%$ branching ratio (BR), into  a pure photino,
which was assumed to be the lightest supersymmetric particle ( LSP ).
 By virtue of this assumption apparently clean limits, independent of
 other SUSY parameters, were obtained on $\GLUM$ and $\MSQ$. 

 Subsequently it was  pointed out \cite{R8} that the squarks and the gluino 
may also decay into other combinations of electroweak  gauginos 
with significantly large  BRs. These Gauginos 
 eventually decay into the LSP which  in general may  not be a pure photino.
Moreover the  $\ETMIS$  spectrum of the LSP in this case is
 much softer.   New analyses \cite {R2} in this more general framework 
revealed  that the limits on  $\MSQ$ and $\GLUM$ were in fact 
less stringent than the earlier ones. Moreover these limits have some (weak) dependence on other SUSY parameters.  

However, the present limits are also not free from  assumptions. They
are based on the predictions of the so called N=1 supergravity ( SUGRA ) 
motivated models \cite{R9} with common scalar and gaugino masses at a 
high scale.
An important byproduct of this assumption is that for a large region of the 
parameter space all scalar superpartners ( including sleptons and sneutrinos ) of the ordinary fermions turn out to
be heavier than the unstable electroweak gauginos into which the squarks and 
gluinos decay. This in turn leads to specific
decay signatures of these sparticles ( see below for the details)
on which the current search strategies are based.
 
 The above assumption though attractive is  by no means compelling.
 The usual assumption that the MSSM is embedded into some Grand
Unified Theory (GUT) usually implies, irrespective of the choice
of any particular gauge group for the GUT, that all gauginos present 
in the model have a common mass at the GUT scale. The assumption of a common gaugino mass at the GUT scale is, therefore, largely model independent
and like most of the  phenomenological  works on SUSY, we shall adopt this. 
 
 On the other hand the assumption of a common scalar mass 
  at the GUT scale is less
 general. In fact many interesting works have shown in recent times 
 that even within the SUGRA frame work one can naturally accommodate
 nonuniversal scalar masses \cite{R10}. This nonuniversality may arise
 simply due the fact that SUGRA may generate common scalar masses
 at the Planck scale. The usual renormalisation group evoution from the
 Planck scale down to the GUT scale may then generate unequal scalar masses 
 depending on the representations to which the scalars belong \cite{R11}.
  Further evolution of these masses down to the electroweak scale 
 may lead to mass patterns significantly different from the conventional ones. There could be more involved mechanisms for  nonuniversal scalar masses at the GUT scale. 
 For example, such nonuniversality can manifest itself
 through D terms \cite{R12} when the GUT group breaks down to
 smaller gauge groups at lower energies. This nonuniversality is quite
 model dependent and depends also on the symmetry breaking chain.  
Moreover in recent literature the possibility of gauge mediated super
symmetry breaking has been discussed extensively \cite{R13}.
Nonuniversality arises in this scenario naturally since the SUSY breaking
mass terms depends on the standard model  quantum numbers of the scalars.

Since it will be a tremendous loss to miss a SUSY signal due to assumptions
inherent in the search strategies, we find it prudent to treat the scalar
masses in a more phenomenological way. This is particularly justified
in view of the above uncertainties in the scalar masses.
We emphasize that certain mass patterns rather than specific choices
for individual sparticle masses, lead to signatures quite different from the conventional ones. Towards the end of this paper we shall discuss the 
compatibility of these patterns with SUGRA motivated theories \cite{R9,R11,R12}. To be specific we work within the following  framework.

The MSSM contains four spin-$\frac{1}{2}$ neutral sparticles. They
 are the superpartners of the photon, the $Z$-boson and the
two neutral $CP$-even Higgs bosons. Linear combinations of these four
states, the four neutral gauginos or neutralinos ($\widetilde{N_i}$,
i=1,4), are the physical states. In the currently favoured models,
the lightest neutralino($\LSP$) is assumed to be the LSP \cite{R1}.
Within R - parity conserving models this is usually assumed to be the
only carrier of $\ETMIS$.
Similarly, linear combinations of the superpartners of the $W$-boson
and the charged Higgs boson give two physical charged gauginos or
charginos. In the following only the lighter chargino ($\W1$) will be
of practical consequences. 
With the usual assumption of a common gaugino mass 
at the GUT scale, the masses 
and the couplings of charginos and neutralinos depend only on three
independent parameters. Usually these are taken as $\mu, \tan \beta$
and the gluino mass $\GLUM$ ( which is related to the universal gaugino 
mass through the  standard renormalisation group equations). Here $\mu$ 
is the soft mass parameter for higgsinos and $\tan \beta$ is the
ratio of the vacuum expectation values of  two neutral Higgs 
bosons in the MSSM. 

If no further assumption is made then the masses of the scalars are
totally independent of the gaugino-masses .
Thus the sneutrinos ($\SNU$, the superpartners of the neutrinos),
though heavier than the LSP, could very well be  lighter than the
$\W1$, the second lightest neutralino ($\N2$) and
other sparticles. As a consequence, the invisible two-body decay
mode $\SNU \longrightarrow \nu \LSP$ opens up and completely
dominates over the others, being the only kinematically-allowed
two-body decay channel for the sneutrinos. The other necessary
condition for this scheme to work is that the $\LSP$ has a
substantial zino (superpartner of the $Z$-boson) component.  This,
however, is almost always the case as long as $\GLUM$ is in
the range interesting for the SUSY searches at the Tevatron
\cite{R2}. Thus the $\SNU$s, decaying primarily into an invisible
channel, may act as additional sources of $\not \!\! E_T$ and can
significantly affect the strategies for SUSY searches \cite{R14,R15}.
It is, therefore, called {\it virtual} or {\it effective} LSP ( VLSP
 or ELSP) \cite{R14,R16} in the context of SUSY searches. 
 Within this basic framework one can think of two closely related 
VLSP scenarios with interesting phenomenological consequences.

I) In some regions of the parameter space right and left sleptons ($\SELR$
and $\SELL$ )along with  sneutrinos may
be lighter than all elctroweak gauginos except  the LSP. By virtue
of the SU(2) breaking mass splitting due to D terms
\begin{eqnarray}
           \MSELL &=&\sqrt{\MSNU^2 + 0.77 D}\\
	    D     &=& - M_Z^2 cos 2 \beta ,
\end{eqnarray}	    

the sneutrinos are always  lighter than the left sleptons. We emphasize
that the above relation holds in a model independent way as long as 
$SU(2)_L$ is a good symmetry above the electroweak scale. Thus irrespective
of the masses of the right sleptons , the sneutrinos will decay into the 
above invisible channel.  By virtue of the assumed mass hierarchy the
 unstable gauginos will almost exclusively  decay into sleptons or sneutrinos
 along with leptons and neutrinos 
via two body decay channels. The sleptons in turn will decay into leptons 
and the LSP. Thus  decays of the electroweak gauginos into jets, 
 which can occur only through three body modes, will be severely suppressed.
 Moreover the number of final states involving one or more leptons  will
 increase.
In contrast in the conventional scenario these  gauginos dominantly decay into jets while only a small fraction of them decay into the leptonic channel.
 As we will see
below, these will have  nontrivial implications for the existing limits
on $\MSQ$ and $\GLUM$ obtained through the jets + $\ETMIS$ channel.

II) In a more restricted region of the parameter space, the $\N2$ --- which 
also has a dominant zino component --- may decay almost entirely  through the  invisible channel $\N2 \longrightarrow \nu \SNU$.
 This happens if both $\tilde{l_L}$ and $\tilde{l_R}$
 are  heavier than the $\N2$. In this special case one obtains two VLSPs . 
Moreover,  $\W1$, usually having  a mass close to that of $\N2$,
  will decay into leptons and $\SNU$s with almost 100 \% BR. As in the previous case the final state will contain smaller number of jets and larger
  number of leptons compared to the conventional scenario.

Some consequences of  both scenario I) and II) (as opposed to the
conventional MSSM where the LSP is the only source of $\ETMIS$)
in the context of SUSY searches at both hadron and $e^+ \, e^-$
colliders  have been discussed in the literature
\cite{R14,R15,R16,R17,R18}. In particular it was qualitatively argued 
in \cite{R14}, that the then existing limits on the $\SQ$ and the $\GLU$
masses from the Tevatron collider, could be relaxed in this scenario.
But no quantitative estimate was given. In this note
we will critically rexamine the current D0 limits \cite{R2}.

 Let us rexamine the D0 \cite{R2} limit $\MSQ > $ 170 GeV,
 for $\GLUM$ = 300 GeV ( for $\mu $= -250 GeV and tan $\beta$ =2 ).
 In this case the SUSY signal is almost entirely controlled
by squark pair production. The produced squarks decay primarily into 
quarks and various electroweak gauginos. In the conventional scenario
the unstable gauginos decay dominantly into jets and the LSP ($\ETMIS$).
Only a relatively small fraction decay into the leptonic channel.
These quarks along with the quark coming from the primary squark decay
contribute to the signal which is required to have at least three hard 
jets each having $E_T >$ 25 GeV \cite{R2}. In the VLSP scenario 
on the other hand
the gauginos decay almost entirely into leptons and / or  $\ETMIS$. Thus
at the parton level each squark pair event will be associated with two
jets only . Of course some events may still have three or more jets due to  fragmentation of the quarks into more than one jet and / or
hard gluon radiations. Yet it is fair to conclude that the three jet 
cuts  affects the signal severely and the existing limit may 
require revision.

The second point of interest stems from the fact that all the $\W1$ s 
and, depending on the slepton masses, some  of the $\N2$ s produced from
squark decays,  eventually decay into leptons. The signal will therefore
be more severely affected by the lepton veto 
than in the conventional case. Of course the
effect of the lepton veto will depend on the $E_T$ of the final state leptons
which in turn depends on the masses  of the gauginos on the one hand  and
that of the sneutrinos and  the sleptons on the other. This point will be 
discussed in further details in the following.

Finally the $\ETMIS$ spectrum is also of considerable interest. Since in this 
scenario  a substantial part of the $N_2 $s decay directly into an invisible mode and the $\W1$ s decay into a carrier of $\ETMIS$ heavier than the LSP via a two body mode, the pattern of missing energy can in principle be diffrent
from the conventinal case.

In Fig.1 we plot the number of
jets (solid lines) in a sample of 10000 events generated by ISAJET 
- ISASUSY \cite{R19} in the conventional(CONV) scenario defined by the 
parameter set (all masses and mass parameters are in GeV and cross 
sections are in pb)
 
  i) $\MSQ$ = 170.0 ( where $\MSQ$ is the mass of five species of 
  L and R  squarks assumed to be degenerate in mass 
 ; as in the D0 paper\cite{R2} We do not include the stop quark in our 
analysis),
$\GLUM$ = 300.0, $\mu$ = -250.0, tan $\beta$ = 2.0 , A=0 and,
following D0, all slepton and sneutrino masses are set equal to the squark mass (in practice any choice 
with sleptons and sneutrinos sufficiently heavier than $\CH$ and $N_2$ gives identical results). 

In Fig.1 ( dashed lines) we also present the same distribution for 
the parameter set 
ii) which is identical to i) except that we now introduce  relatively light 
sneutrinos with $m_{\SNU}$ = 50.0 and the L type sleptons with mass given 
by eqn 1. For definiteness we have taken $\MSELL$ = $\MSELR$ but this choice 
has little  bearing on the final results. The relevant electroweak 
gaugino masses for  set i) and ii) are
$\MLSP$ = 48, $\MCH1$ = 101.76 and  $\MN2$ = 102.27.
  
It is clear from Fig.1 that in the conventional scenario
most of the events are associated with three
or more jets. In contrast, the VLSP scenario represented by the 
parameter set ii) shows a peak at $N_{jet}$ = 2 as expected. Thus
a large fraction  of the signal events will be removed by the three jet cut. 

In Fig.2  we compare the $E_T$ spectrum of the softer lepton
( the spectrum of both the leptons in the final state look quite
similar). Again it is quite clear that the lepton veto (defined below) affects
scenario ii) quite severely.

Finally in Fig.3  we compare the $\ETMIS$ spectrum for set i)
( solid lines ) and ii)( dashed lines ). They turn out to be quite similar.

It is now natural to rexamine the limits on $\MSQ$ in scenario ii). While 
 a rigorous new limit cannot be obtained without full
detector simulations, the trend can be easily understood from our
analysis. We apply the following cuts introduced in the D0 paper
to reduce the SM background \cite{R2}:\\
A) We require $\ETMIS$ $> $ 75 GeV.\\
B) The signal is rquired to have three or more jets with $E_T > $ 25 GeV
and $|\eta| < $ 3.5 .\\
C) Jets are ordered in terms of decreasing $E_T$. We then require that 
the leading jet should not fall into the calorimeter crack, i.e., its 
pseudorapidity should not lie in the region 1.1 $ \leq |\eta| 
\leq $ 1.4.\\
D) The angle $\delta \phi_k$ is defined as the  azimuthal angle between 
the kth jet and the $\ETMIS$ vector. Events having $\delta \phi_k > 
\pi$ - 0.1 or
$\delta \phi_k < $ 0.1 for k = 1, 2, 3 are rejected.
 Events with $\sqrt{(\delta \phi_1 - \pi)^2  + (\delta \phi_2)^2 }<$ 0.5 
are also rejected.\\
E) The lepton veto (there will be no electrons with $E_T >$ 20 GeV and muons 
with $E_T > $ 15 GeV) is then applied.\\
In table I we present the  signal cross section ( $\sigma_c$ )
after cuts for  parameter sets i) and ii). The efficiencies of the cuts
A) - E) can also be obtained from this table where the number of generated
events is 10000. 
Since the lepton veto plays a major role in this analysis,
we present results for three values of lepton detection efficiency (LDE) :
100 $\%$, 70 $\%$ and 60 $\%$ ( the three numbers in the 8th and the last
column). It follows from table I that the 
reduction of the cross section in scenario ii) compared to that in  scenario 
i), is substantial even for LDE as low as 60$\%$. It is also clear 
 that the cuts B) and E) reduces the signal more drastically in scenario ii).

In order to have a rough idea about the changes in  the squark mass limit we now compute the cross sections for the following sets of parameters\\
iii) which is the same as in ii) except that $\MSQ$ = 150,\\
iv) which is the same as  in ii) but with $\MSQ$ = 160.\\
The results are presented in tables I . We find that for low LDE both 
choice iii) and iv)  give cross sections  consistent with the one obtained 
for  set i) which represents the published  95 \% CL D0 limit on
$\MSQ$, corresponding to a candidate sample of 14 events in  
13.5 $\pm$ 1.6 pb$^{-1}$ data, against an estimated background of 17.1 $\pm$ 
1.8 $\pm^{7.0}_{6.6}$\cite{R2}. Thus the current limits on $\MSQ$ for $\GLUM$ = 300.0, 
is likely to be reduced by 10 - 20 GeV
in the VLSP scenario. 

At this point it might be interesting to examine the sensitivity of the signal 
to $\MSNU$. A clear advantantage of SUSY search in the jets + $\ETMIS$ channel 
will emerge from this discussion. In the VLSP scenario the unstable gauginos
primarily decay into leptons giving enhanced signal in the leptonic channel 
compared to the conventional scenario. For example, SUSY signals  from
chargino pair production followed by their leptonic decays at both $e^+ - e^- $ and hadron colliders \cite{R16,R18} in the VLSP scenario have been discussed in the literature. The typical signal is a pair of hadronically quiet dileptons which can be 
 seperated from the aparently large W-W , Drell - Yan or two photon
backgrounds by suitable kinematical cuts. However,
in all these cases the resulting signal becomes weak if the mass difference
between the chargino and the sneutrino is  small.  The reason
is rather obvious. With heavier sneutrinos the signal leptons become softer
and becomes indistinguishable from the Drell - Yan or two photon back ground.
Similarly,  the 2l (dilepton ) + jets + $\ETMIS$ signal is also significantly 
enhanced in the VLSP scenario compared to the conventional
case\cite{R14,R15}. As we shall show below this signal also becomes weaker
and no useful limit can be derived, if the above mass difference  
happens to be small.

On the other hand the limit that follows from the jets + $\ETMIS$
channel for a large chargino - sneutrino mass difference is rather 
conservative . This is because for heavier sneutrinos the lepton veto 
affects the signal to a lesser extent and the squark mass limit become
stronger. This is illustrated in table I for the parameter set\\
v) which is the same as in iii) except that $\MSNU$ = 92. \\
With chargino masses as stated 
above the mass difference between the chargino and the sneutrino is about 10.
The resulting enhancement of the cross section compared to  set choice iii)
is to be noted. We, however, emphasize that even in this case the 
reduction of the signal due to the 3 jet cut takes place. The signal, 
therefore, looks considerably different from the conventional one. We have
also noted that the signal remains practically unaffected by 
specific choices of $\MSNU$, as long as 
the chargino sneutrino mass difference remain large. This has beeen
verified by  varying $\MSNU$ in the interval 50 - 75 GeV for the parameter set iii). For example with $\MSNU$ = 75 GeV, we obtain $\sigma_c$
= 2.01, 2.11 and 2.24 for three LDEs which is certainly consistent with
the result for set iii) ( see table I ).

Although the above analysis is for a specific $\MSQ$ and $\GLUM$, we wish
to stress that the current D0 limits on $\MSQ$, for $\GLUM  > \MSQ$, is
likely to be affected by the VLSP scenario if $\GLUM <$ 400 GeV. For
$\GLUM$ in this range, squarks having masses within the 
current D0 limits are kinematically allowed to decay into $\W1$ and $\N2$.
Due to  hadronically quiet  decays of these gauginos in the VLSP scenario
as discussed above, relaxation of these limits is to be expected. For higher $\GLUM$, however, the limits on $\MSQ$ are rather weak and
both the $\W1$ and the $\N2$ happen to be heavier than 
 squarks having masses consistent with the current limits. These
 squarks decay directly into LSP with $100 \%$ BR. Thus the signal is  similar both in the conventional and VLSP scenarios.
 As a result further relaxation of the existing limits is
improbable, although as and when these limits get stronger VLSP effects
should be taken into account.

We now turn our attention to the mass limits when the squarks and the
gluinos are approximately degenerate in mass. The existing limit is
$\MSQ \simeq \GLUM >$ 212 GeV. In table II we present the cross section for
the set of parameters with the kinematical cuts A) - E)\\
vi) $\MSQ \simeq \GLUM \simeq $ 212. All other parameters as 
   in set i).\\
   vii) $\MSQ \simeq \GLUM \simeq $ 202, $\MSNU$ = 50 and $\MSELL$ as 
   determined by eqn 1. For simplicity we also take $\MSELL$ = $\MSELR$,
although, as discussed earlier this choice has little bearing on
 the final result. Other SUSY parameters are chosen as above.\\
    viii)  $\MSQ \simeq \GLUM \simeq $ 205 with other parameters as in vii).

Comparing the results in table II it is fair to conclude that a reduction
of the existing limits by 7 - 10 GeV seems to be very likely. Although this 
relaxation is modest compared to  that in the previous example, this set
of $\MSQ$ and $\GLUM$ deserves special attention. For this set a relatively
light sneutrino can be accommodated more easily in  SUGRA motivated frameworks
\cite{R9,R11} as will be discused below.

For $\MSQ >> \GLUM $ , the existing limit  $\GLUM >$ 146 GeV is rather weak. For this 
$\GLUM$, $\MCH1 \simeq$ 46 GeV. Thus there is very little parameter space 
available for a light sneutrino. Any relaxation of the existing limit is therefore unlikely. However, as more data accumulates a significant
improvement of this limit is expected. When that happens 
the VLSP scenario will have important bearings on the limits. This is 
because the parameter space available in the VLSP scenario
in the $\GLUM$ - $\MSNU$ plane increases with $\GLUM$ \cite{R18}.

 We now turn our attention to the 2l + jets + $\ETMIS$ channel. As has
already been discussed \cite{R14,R15} the enhanced leptonic branching 
ratio of the
unstable electrtoweak gauginos in the VLSP scenario is likely to give 
a larger signal in this channel. On the other hand the stringent 
cuts on the  
leptons required to suppress the SM background, make the signal very 
sensitive to the mass differnce between the gauginos and the scalars.
In particular the mass difference between the lighter 
chargino and the sneutrino plays a  crucial role as we shall presently 
see. The Following kinematical cuts are imposed:\\
F) The signal is required to have at least two electrons with $E_T >$
15 GeV for $|\eta |< $ 2.5.\\
G) It  is further required that there be at least two jets with $E_T >$
20 GeV for $|\eta |< $ 2.5.\\
H) $\ETMIS$ is required to be $>$ 25 Gev.\\
I) The invariant mass of the electrons $m_{ee}$ is required to satisfy
   $|m_{ee} - M_Z| > $ 12 GeV. If this is not satisfied then $\ETMIS$
   is required to be $>$ 40 GeV. \\

In table III we present the effects of the above cuts on 10000 events 
generated by ISASUSY - ISAJET \cite{R19} for the parameter sets i), ii) and\\
  ix) all parameters as in ii) but $\MSNU$ = 92 GeV.

We have conservatively assumed only 30\% efficiency for detecting two 
electrons. It follows from table III that in contrast to the jets + $\ETMIS$
channel, set ii) indeed gives a larger signal in the dilepton channel.
When data for an integrated luminosity of $\approx$ 100 pb $^{-1}$ is available, the VLSP scenario can indeed be probed in this channel. In this
analysis we have restricted ourselves 
 to electrons only. This is because the muon detection efficiency is poor for the D0 experiment. Even  the CDF experiment has a rather
limited angular coverage for muon detection. The physics of the search
in the dimuon channel or e-$\mu$ channel in the VLSP scenario is, however,
very similar. Inclusion of muons along with improvements in LDE  will 
make SUSY search in the dilepton channel more attractive in the VLSP 
scenario.
 
As mentioned earlier the improved signal in the dilepton channel is 
expected to be strongly dependent on the sneutrino - chargino mass difference and is likely to become extremely weak when this difference is $\simeq$ 10 GeV. This is illustrated in table III with parameter set ix) .       	  

It may be noted that the CDF collaboration has recently published limits
on squark - gluiono masses based on searches in the dilepton channel
\cite{R20}. It is however not possible to compare our results with theirs,
since they used next to leading order cross sections for squark -
gluino production \cite{R21}. 

We now turn our attention to the possibility of accommodating the
VLSP scenario in   N=1  SUGRA models with a commom scalar mass 
at the GUT scale. A similar analysis was made in \cite{R16} but
the squark gluino mass limits from TEVATRON were not taken into account 
 since the impact of the VLSP scenario on these limits were 
not known.  The solution of the standard RG equations does not give
identical masses for L and R squarks of different flavours although,
 apart from the top squarks, the masses are closely spaced. On the
other hand, the experimental limits  are given in terms of a common
 squark mass. We identify this  with the average   mass of two up  
 and three down type squarks of both L and R types.
 Using the standard results \cite{R9} we obtain: 
\begin{equation}
  \MSQ^2 = m_0^2 + \frac {8}{9} \GLUM^2 - 0.63 M^2 - 0.05 D  
\end{equation}
where $m_0$ and M are respectively the common scalar and gaugino mass
at the GUT scale and the constant D has been defined earlier. In order
to reduce numerical uncertainties  M dependent terms have been written 
directly in terms of $\GLUM$ wherever possible. Substituting for $m_0^2$ in the standard expression for $\MSNU^2$, we obtain
\begin{equation}
  \MSNU^2 = \MSQ^2  - \frac {8}{9} \GLUM^2  + 1.15 M^2 + 0.55 D .  
\end{equation}
M is related to $\GLUM$ by the relation 
\begin {equation}
 M = \GLUM  ( 1 + \beta_3 t)
\end{equation}
where  $\beta_3$ = $ \frac{ -3 \alpha_0}{ 4 \pi} $ and t = ln $\frac
{M_G^2}{Q^2} \approx$  66.0, $M_G$ being the GUT scale and Q the appropriate
low energy scale.  The value of $\alpha_0$, the SU(5) fine structure 
constant at $M_G$,  is uncertain mainly due
 the present uncertainty in the value of the strong coupling constant.
 A recent fit \cite{R22} gives $\alpha_0^{-1}$ = 23.9 - 25.3. It can
 be readily checked from the above equation that due to the above 
uncertainty
$\MSNU$ can vary considerably for  given $\MSQ$ and $\GLUM$. For example,
with $\MSQ$ = 200 GeV, $\GLUM$ = 215 GeV and $\alpha_0$ in the above range,
$\MSNU$ = 48 - 62 GeV. In the entire range $\MSNU < \MCH1$ and , hence 
sneutrinos decay  invisibly. It is worthwhile to note that for $\GLUM$
= 215 GeV, the above  $\MSQ$ is indeed close to the present D0  limit.
Hence a revaluation  of this limit is indeed called for.
For definiteness we shall use in the following $\alpha_0^{-1}$ = 24.3 which
is a typical fit value of \cite{R22}.

If the idea of a common scalar mass at the GUT scale is taken seriously,
then the following conclusions follow from the above equations. 
\begin{itemize}
\item The present limit $\MSQ \approx \GLUM$ = 212 GeV is not affected by the
VLSP scenario, since in this case $\MSNU \approx $ 92 GeV and is larger than
$\MCH1$. The same conclusion holds for $\MSQ$ limits for lower $\GLUM$.
\item For larger $\GLUM$ ,however, all squark masses are not allowed as they
may lead to unacceptably low or even unphysical values for $\MSNU$ ( the 
bound on $\MSELR$ from LEP 1.5 should also be kept in mind; this bound of 
course is the
same in both conventional and VLSP scenarios). For example using the limit
$\MSNU >$ 45 GeV, one readily obtains for $\GLUM$ = 220 GeV , $\MSQ > $
202 GeV. On the other hand
for $\MSQ \approx$ 216 GeV , $\MSNU \approx \MCH1$, which in this case 
 is $\approx$ 86 GeV. Thus in the narrow but
interesting range 202 GeV$ \leq\MSQ \leq $ 216, the effect of the VLSP
scenario should be taken into account. Similar ranges can be obtained for
higher values of $\GLUM$.

\end{itemize}
As has been emphasized at the beginning of this paper, the idea of a common
scalar mass at the GUT scale may be subjected to revision.
 For example if one  assumes that all the scalars have a common mass
at the Planck scale \cite{R11}, which certainly is reasonable,
 then  revison of eqn 3) and 4) is called for.  Within the frame work of a
SU(5) GUT, the $\tilde{u}_L, \tilde{d}_L, \tilde{u}_R$ and $\tilde{l}_R$
belong to the 10 of SU(5). Their masses when evolved down to the GUT scale
follow a certain pattern. On the otherhand $\SELL, \SNU$ and $\tilde{d}_R$
belong to the $\bar{5}$ of SU(5). As a result their masses evolve differently.
Using eqn 9 of \cite{R11}, it is straightforward to show that the 
 modified equations are :
\begin{equation}
  \MSQ^2 = m_0^2 + \frac {8.8}{9} \GLUM^2 - 0.22 M^2 - 0.05 D  
\end{equation}
\begin{equation}
  \MSNU^2 = \MSQ^2  - \frac {8.8}{9} \GLUM^2  + 1.07 M^2 + 0.55 D  
\end{equation}
It is now follows that more regions of the parameter space opens up
for the VLSP scenario. For example, with $\MSQ \approx \GLUM \approx$ 
205 GeV,
$\MSNU$ = 61 GeV. Thus the relaxed limit discussed above is certainly
compatible to a SUGRA motivated scenario. The VLSP scenario can now be 
accommodated, for almost all $\GLUM$s within the striking range of 
Tevatron, for $\GLUM \approx \MSQ$. As shown above, for $\GLUM > 
\MSQ $,  the VLSP scenario can be accommodated for a range of $\MSQ$.
   
 If the GUT group is SO(10), then all  squarks and sleptons belong to
 the  16 plet. In this case even if a common scalar mass is generated
 at the Planck scale, the scalars will continue to have the same mass
 at the GUT scale. However, when SO(10) breaks down to groups of lower 
 rank then as discussed in \cite{R12}, nonuniversality of scalar masses
 may arise due to D term contributions. This will always happen if some of 
 the heavy fields which are integrated out have nonuniversal SUSY breaking        masses. For example if SO( 10 ) directly breaks down to the SM,
  the initial conditions for diffrent scalar masses
 at $M_G$ can be parametrised as follows \cite{R12}. For the $\tilde{u}_L
, \tilde{d}_L, \tilde{u}_R, \tilde{l}_R$ we have at $M_G$

\begin{equation}
    m_0^2 = m_{16}^2 + \delta m^2
\end{equation}
On the other hand the common mass for the $ \tilde{d}_R, \tilde{l}_L$ and 
$\tilde{\nu}_L$ is given by
\begin{equation}
    m_0^2 = m_{16}^2 - 3 \delta m^2
\end{equation}
where $m_{16}$ is the common scalar mass at $M_G$ and $\delta m^2$, the
D term contribution, is essentially a free parameter.
The reulting eqns for the average squark mass squared and $\MSNU^2$ are
modified to 
\begin{equation}
  \MSQ^2 = m_0^2 + \frac {8}{9} \GLUM^2 - 0.63 M^2 - 0.05 D - 0.2 \delta m^2 
\end{equation}
and
\begin{equation}
\MSNU^2 = \MSQ^2  - \frac {8}{9} \GLUM^2  + 1.15 M^2 + 0.55 D - 2.8 \delta m^2
\end{equation}
For $\MSQ \approx \GLUM \approx $ 205 GeV, one obtains $\MSNU \approx$ 50 GeV
with $\delta m^2 \approx$ 1900 GeV$^2$. Thus a light sneutrino can be easily
accommodated. However, for this $\delta m^2$ the $\tilde{d}_R$ type squarks
become considerably lighter than the other squarks. For example the mass 
difference between  $\tilde{u}_R$ and  $\tilde{d}_R$ type squarks become
$\approx$ 20 GeV. Thus one may question the idea of average squark mass.
Perhaps a better parametrisation in this case will be to take an average
mass for  $\tilde{u}_L$  $\tilde{d}_L$ and $\tilde{u}_R$ squarks and a somewhat lower mass for the  $\tilde{d}_R$ squarks. Such a scenario would
then naturally lead to a relatively light sneutrino. Sparticle mass limits
in this scenario will indeed be interesting.

In conclusion we reiterate that if the sneutrino is indeed the second 
lightest SUSY particle, then the decay signatures of the squarks and 
the gluinos become  considerably different from the conventional ones. This 
may relax the existing limits on $\MSQ$ and $\GLUM$ obtained in the jets + 
$\ETMIS$ channel. More importantly, the light sneutrino is likely to play
a more prominent role as the limits get stronger with the accumulation of
data. As noted earlier \cite{R14,R15}, the signal in the 2l + jets
+$\ETMIS$ channel is enhanced while the signal in the trilepton
+ $\ETMIS$ channel either gets depleted or become weaker in this 
scenario. Such a light sneutrino arise quite naturally in various
SUGRA motivated frame works.\\ 
\noindent
{\bf Acknowledgement}: AD thanks 
the International Centre for Theoretical
Physics, Trieste, Italy for hospitality. AD's work was partially
supported by the Department of Science and Technology, Govt. of India.
The authors gratefully acknowledge discussions with D. P. Roy,
M. Bisset and N. K. Mondal. MG and NP wish to give thanks to Subir Sarkar 
and Suchandra Dutta for helping in computation.



\newpage

\begin{flushleft}
{\Large \bf Table Captions}\\
\end{flushleft}

\vskip 10pt
\begin{flushleft}
{\bf Table-I :}
The number of events for parameter sets i)-v) have been 
shown after the cuts  A - E (see text). 
 $\sigma_{p}$ is the production cross section 
of gluino and/ or squark pairs, $\sigma_{c}$ is the signal cross section
after cuts. The three numbers in  column E and in the last column 
correspond to 100\%, 70\% and 60\% LDE. 
\end{flushleft}

\vskip 12pt
\begin{flushleft}
{\bf Table-II :}
The gluino-squark pair production ($\sigma_{p}$)
and signal cross sections ($\sigma_{c}$) for  parameter 
sets (vi) - (viii) see text are shown. 
\end{flushleft}

\vskip 12pt
\begin{flushleft}
{\bf Table-III :}
 The number of dilepton events  after cuts F - I are shown
for parameter  sets (i),(ii) and (ix) (see text).   

\end{flushleft}

\section{Figure Captions}
\renewcommand{\labelenumi}{Fig. \arabic{enumi}}
\begin{enumerate}

\vspace{10mm}
\item   
\noindent
{\small
The distribution of jets with the number of events. 
The solid(dashed) lines correspond to the parameter set(i)((ii)).} 

\item  
\noindent
{\small
The distribution of $E_{T}$ of soft leptons  with the number of events
following the conventions of Fig. 1.}  
\item 
\noindent
{\small
The distribution of $\ETMIS$  with the number of events following the conventions of Fig 1.}
\end{enumerate}

\newpage


\vskip 20pt
\begin{center}
{\bf Table-I}
\end{center}
\begin{center}
\begin{tabular}{|c|c|c|c|c|c|c|c|c|c|}
\hline
Set&$\GLUM$  $\MSQ$ & $\MSNU$  $m_{\tilde l}$ &  A & B & C & D & E & $\sigma_{p}$ & $\sigma_{c}$\\
\hline
i&300  170 &170  170 &  4553 & 2114 & 1969 & 1541 & 1284 & 16.5 & 2.12 \\
 &         &         &       &      &      &       & 1361 &      & 2.24 \\
 &         &         &       &      &      &      & 1387 &      & 2.29 \\  
\hline
ii&300 170 & 50 79 &  4824 & 1750 & 1618 & 1250 & 830 & 16.6 & 1.37 \\
  &        &       &       &      &      &      &  956 &     & 1.58 \\
  &        &       &       &      &      &      &  998 &     & 1.65 \\ 
\hline
iii&300 150 & 50 79 & 3937 & 1264 &1175 & 914 & 624  & 31   & 1.93 \\
  &  &      &      &       & &     & 711  &      & 2.20 \\
  &  &      &      &       &  &    & 740  &      & 2.29 \\ 
\hline
iv&300 160 & 50 79 &  4512& 1602 & 1492 & 1163 & 790  & 23   & 1.87 \\
  &  &      &      &       & &     & 900  &      & 2.07 \\
  &  &      &      &    &    &  & 939  &      & 2.15 \\ 
\hline
v&300 150 & 92 111&  3742 & 1087 & 1015 & 805 &  743  & 31   & 2.3  \\
 &   &      &      &      & &     & 762  &      & 2.36 \\
 &   &      &      &      &  &    & 768  &      & 2.38 \\ 
\hline
\end{tabular}
\end{center}

\vskip 5pt
\begin{center}
{\bf Table-II}
\end{center}
\begin{center}
\begin{tabular}{|c|c|c|c|c|}
\hline
Set & $\GLUM$  $\MSQ$ & $\MSNU$  $m_{\tilde l}$&$\sigma_{p}$ & $\sigma_{c}$\\
\hline
vi&212 212&212 212 & 10.9 & 2.15 \\
  &   &  &      &  2.30 \\
  &    &  &      &  2.35 \\
\hline
vii&202 202 & 50 79 & 15.8 & 2.24 \\
   &  & &        & 2.48 \\
  &   & &        & 2.57 \\
\hline
viii&205 205 & 50 79& 14.2 & 2.09 \\
  & &  &         & 2.33 \\
  & &  &         & 2.41 \\
\hline
\end{tabular}
\end{center}
\newpage
\vskip 20pt
\begin{center}
{\bf Table-III}
\end{center}
\begin{center}
\begin{tabular}{|c|c|c|c|c|c|c|}
\hline
Set&$\GLUM$  $\MSQ$ & $\MSNU$  $m_{\tilde l}$ & F & G & H & I\\
\hline
i &300 170 &170 170 & 243 & 217 & 190 & 190 \\
\hline
ii &300 170 &50 79 & 287 & 253 & 232 & 228 \\
\hline
ix&300 170 &92 111& 27 & 21 & 19 & 19 \\
\hline
\end{tabular}
\end{center}
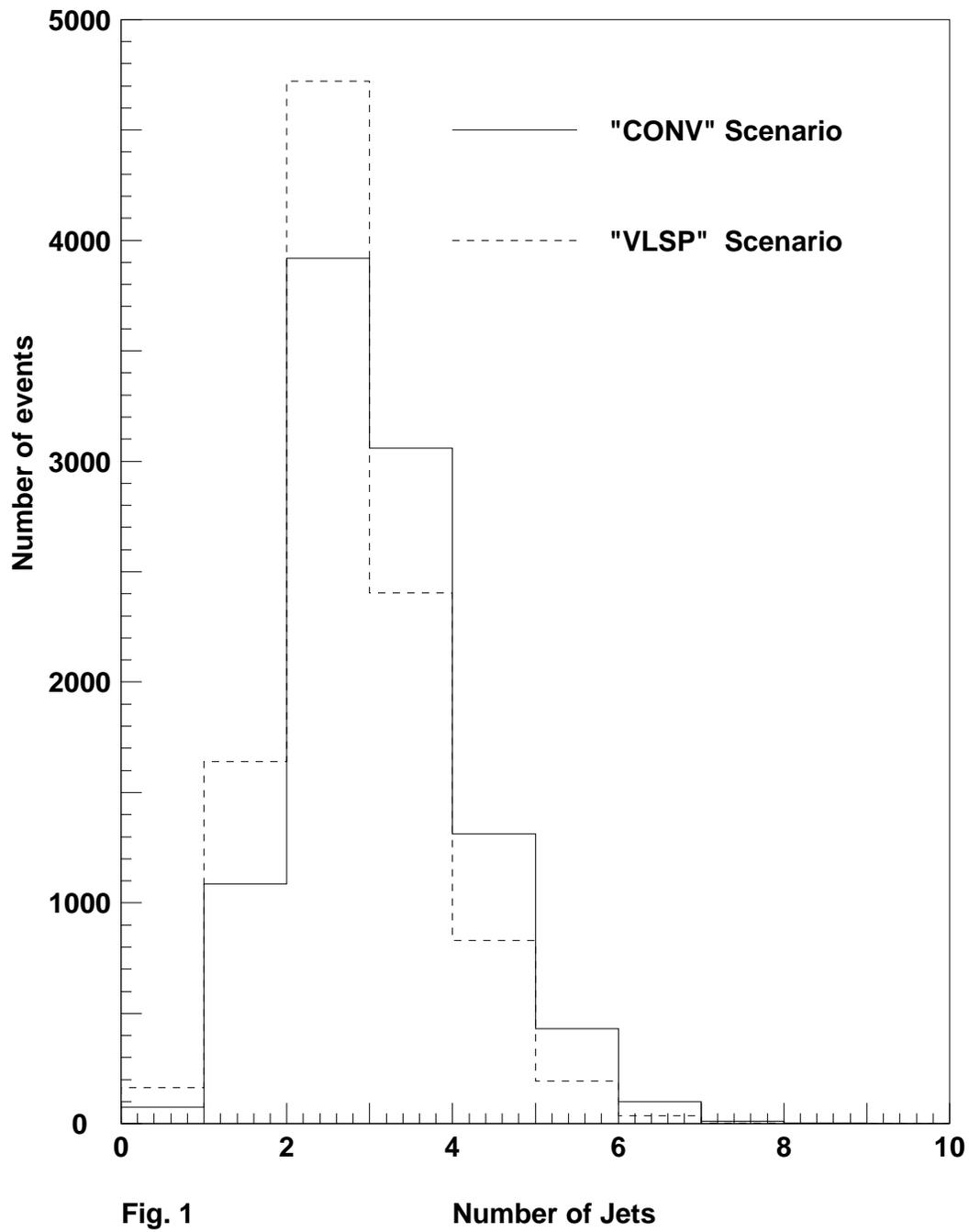
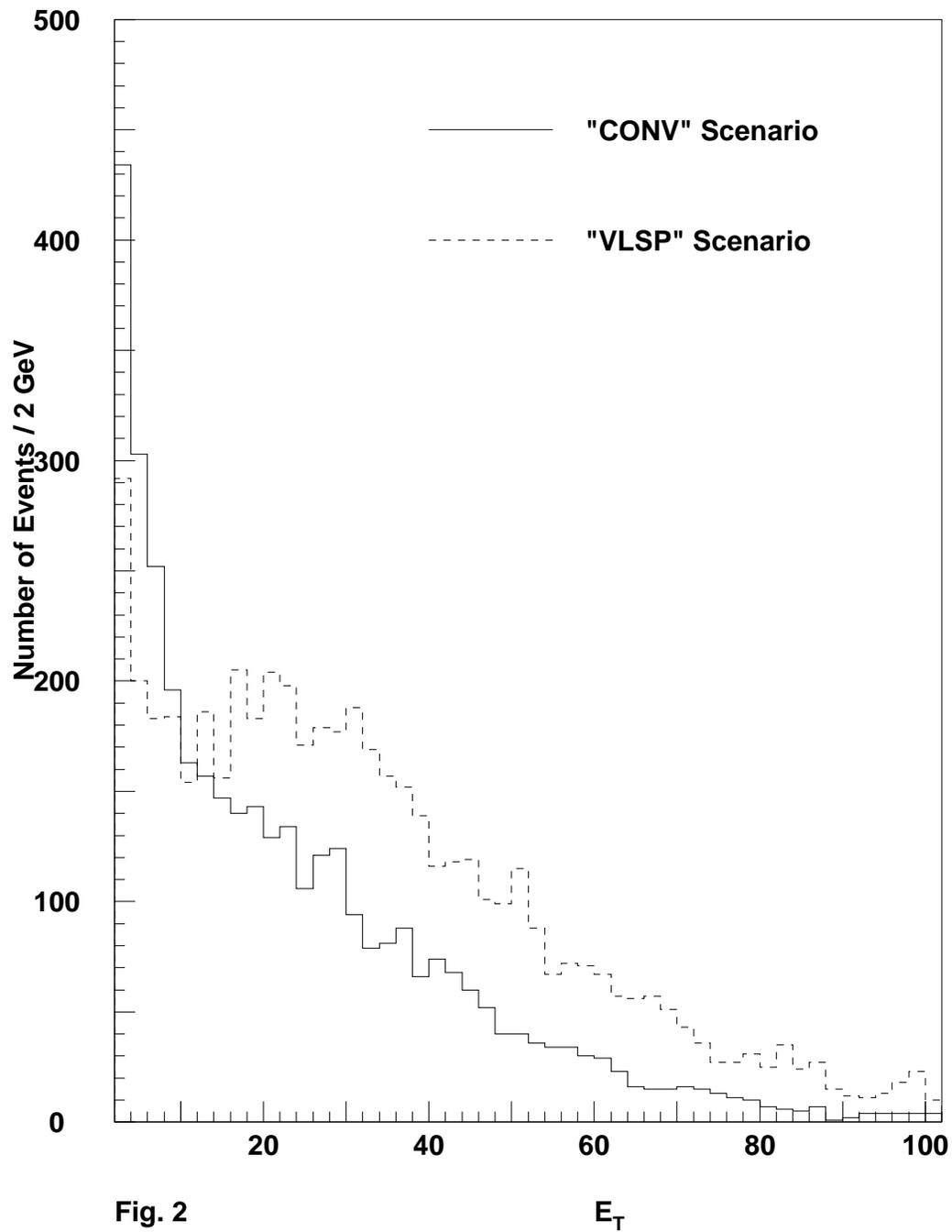
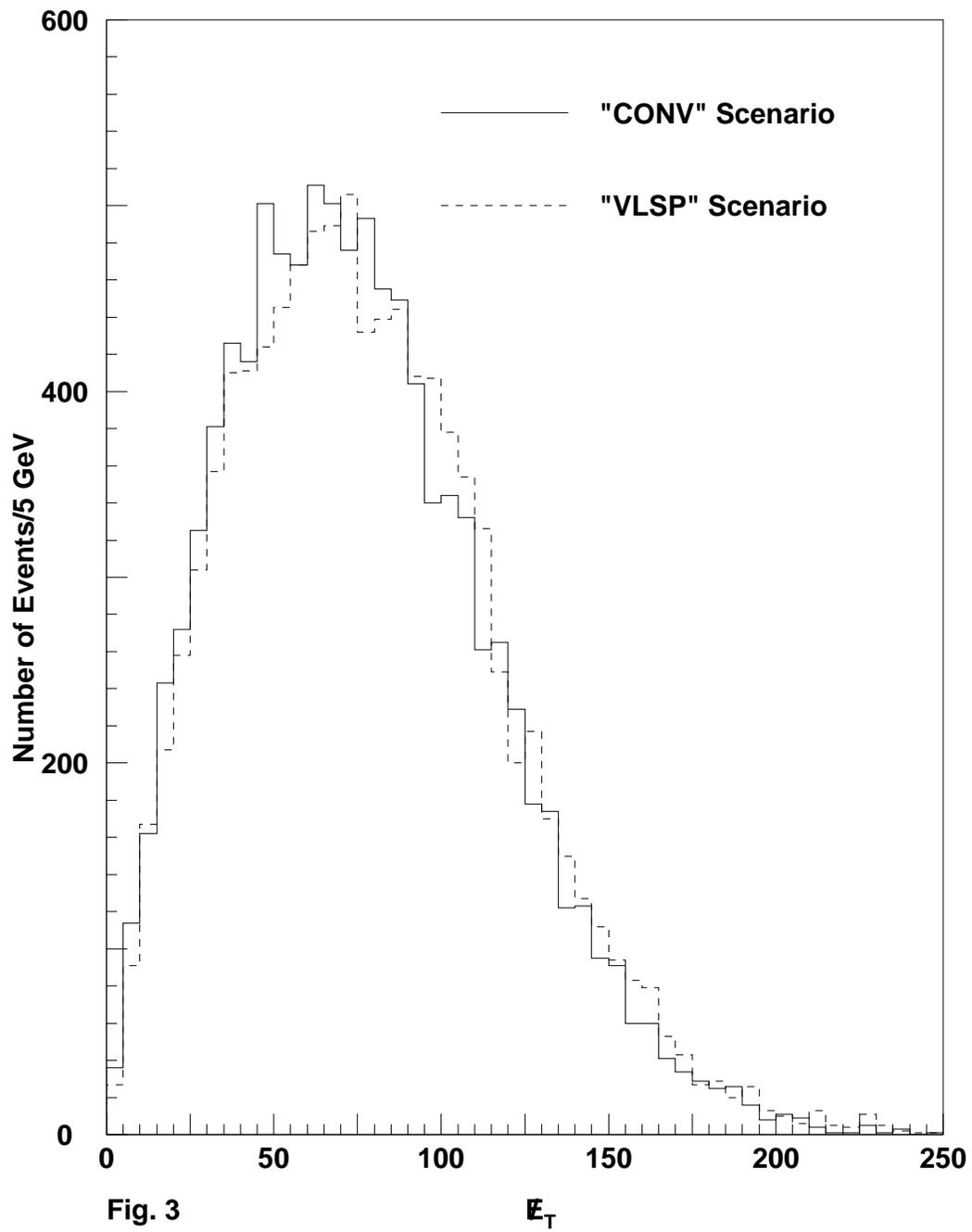
\end{document}